\documentclass{appolb}
\usepackage{epsfig}
\usepackage{axodraw}

\newcommand{\be}{\begin{equation}}
\newcommand{\ee}{\end{equation}}
\newcommand{\ba}{\begin{eqnarray}}
\newcommand{\ea}{\end{eqnarray}}
\newcommand{\dis}{\displaystyle}

\begin{document}

\begin{titlepage}
\begin{flushright}
CERN-PH-TH/2007-020\\
LU TP 07-02\\
\end{flushright}
\vspace{2cm}
\begin{center}

{\large\bf Status of the Hadronic Light-by-Light Contribution to the
 Muon Anomalous Magnetic Moment
\footnote{Talk given by J.P.
at ``Final Euridice Meeting'', August 24-28  2006, Kazimierz, Poland.}}\\
\vfill
{\bf  Johan Bijnens$^{a)}$
 and Joaquim Prades$^{b)}$
\footnote{On leave of absence from CAFPE and Departamento de
 F\'{\i}sica Te\'orica y del Cosmos, Universidad de Granada, 
Campus de Fuente Nueva, E-18002 Granada, Spain.}
}\\[0.5cm]

$^{a)}$  Department of Theoretical Physics, Lund University\\
S\"olvegatan 14A, S-22362 Lund, Sweden.\\[0.5cm]

$^{b)}$ Theory Unit, Physics Department, CERN \\ 
CH-1211  Gen\`eve  23, Switzerland.\\[0.5cm]

\end{center}
\vfill
\begin{abstract}
\noindent
We review the present status of
the hadronic light-by-light contribution to muon $g-2$
and critically compare recent calculations. 
\end{abstract}
\vfill
January 2007
\end{titlepage}
\setcounter{page}{1}
\setcounter{footnote}{0}

\title{Status of the Hadronic Light-by-Light Contribution to the 
Muon Anomalous Magnetic Moment
\thanks{Presented by JP.}}

\author{Johan Bijnens$^{(a)}$ and Joaquim Prades$^{(b)}$\footnote{On 
leave of
absence from CAFPE and Departamento de
 F\'{\i}sica Te\'orica y del Cosmos, Universidad de Granada, 
Campus de Fuente Nueva, E-18002 Granada, Spain.}
\address{$^{(a)}$Dept. of Theor. Physics, Lund University,
S\"olvegatan 14A, S-22362 Lund, Sweden\\[2mm]
$^{(b)}$Theory Unit, Physics Dept., CERN,
CH-1211  Gen\`eve  23, Switzerland.}
}

\maketitle

\begin{abstract}
We review the present status of
the hadronic light-by-light contribution to muon $g-2$
and critically compare recent calculations. 
\end{abstract}

\section{Introduction}

The  muon anomalous magnetic moment
$g-2$ [$a_\mu \equiv (g-2)/2$]  has been measured by the 
E821 experiment (Muon g-2 Collaboration)
at BNL with an impressive accuracy of 0.72 ppm \cite{BNL06}
yielding the present
world average~\cite{BNL06}
\be
a_\mu^{\rm exp} = 11 \, 659 \, 208.0(6.3) \times 10^{-10} \, 
\ee
with an accuracy of 0.54 ppm.
 New experiments \cite{ROB06,MRR06} are being designed to 
measure $a_\mu$ with an accuracy of at least 0.25 ppm.

On the theory side,  a lot of work has been devoted to reduce the
uncertainty of the Standard Model prediction. For a recent updated
discussion see \cite{MRR06} where an extensive list of references
for both theoretical predictions  and experimental results  
can be found.

\begin{figure}
\label{lbl}
\begin{center}
\epsfig{file=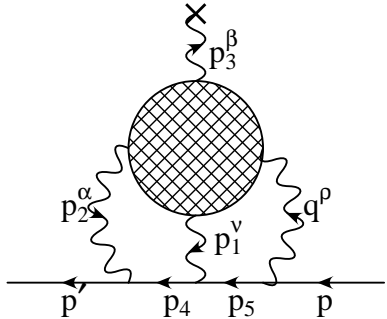,width=5cm}
\end{center}
\caption{The hadronic light-by-light
contribution to the muon $g-2$.}
\end{figure}

Here, we  critically review the present status
of the hadronic light-by-light contribution whose uncertainty will
eventually become the largest theoretical error.
This contribution is depicted  Fig.~\ref{lbl}.
It consists of three photon legs coming from
the  muon line connected to
the external electromagnetic field
by hadronic processes.
Its contribution can be written as
\ba
\label{Mlbl}
\dis{\cal M}
&=& \vert e\vert^7 V_\beta
\int {{\rm d}^4 p_1 \over (2\pi )^4}
\int {{\rm d}^4p_2\over (2\pi )^4} \,
{1\over q^2\, p_1^2 \, p_2^2 (p_4^2-m^2) \,
(p_5^2 - m^2)} \nonumber \\
&\times&  \Pi^{\rho\nu\alpha\beta} (p_1,p_2,p_3) \,
\bar{u}(p^\prime )\gamma_\alpha (\not{\! p}_4 +m )\gamma_\nu
(\not{\! p}_5 +m ) \gamma_\rho u(p) \, 
\ea
where $q=p_1+p_2+p_3$. To get the amplitude ${\cal M}$ in (\ref{Mlbl}), 
one needs the  full correlator $\Pi^{\rho\nu\alpha\beta}(p_1,p_2,p_3 \to 0)$, 
with 
\ba
\label{four}
\Pi^{\rho\nu\alpha\beta}(p_1,p_2,p_3)&=&
i^3 \int {\rm d}^4 x \int {\rm d}^4 y
\int {\rm d}^4 z \, \e^{i (p_1 \cdot x + p_2 \cdot y + p_3 \cdot z)}
\times \nonumber \\
&\times& \langle 0 | T \left[  V^\rho(0) V^\nu(x) V^\alpha(y) V^\beta(z)
\right] |0\rangle 
\ea
with $V^\mu(x)=\left[ \overline q \hat Q \gamma^\mu q \right](x)$
and $\hat Q = {\rm diag}(2,-1,-1)/3$ the quark charges.
The photon leg with momentum $p_3 \to 0 $ couples to
the magnetic field.

Clearly, because we have two fully independent momenta
many different energy scales are involved in the
calculation of the  
hadronic light-by-light contribution to muon $g-2$.
This makes it difficult to obtain the full needed behaviour
of the correlator (\ref{four}) from known constraints.
Therefore no full first principles
 calculation exists at present.
The needed results cannot be directly
related to measurable quantities either.
A first exploratory lattice QCD calculation has been attempted\cite{HBIY06}.

 Using $1/N_c$ and chiral perturbation theory (CHPT)
 expansion counting, one can organize the
 different contributions \cite{EdR94}:
\begin{itemize}
\item Goldstone boson exchange contributions are order $N_c$ and start
contributing at order $p^6$ in CHPT.
\item (Constituent) quark-loop 
and non-Goldstone boson exchange contributions  are order
$N_c$ and start contributing at order $p^8$ in CHPT.
\item Goldstone boson loop contributions are order  one in $1/N_c$
and start contributing at order $p^4$ in CHPT.
\item Non-Goldstone boson loop contributions are order one in $1/N_c$
 and start to contribute at order $p^8$ in CHPT.
\end{itemize}

The two existing {\em full} calculations,  \cite{BPP96}
and  \cite{HK98} are based on this classification.
The Goldstone boson exchange contribution (GBE) was shown to be numerically
dominant in \cite{BPP96,HK98} after strong cancellations between the other 
contributions. \cite{KNPR02} showed that the leading double logarithm comes
from the GBE and was positive. \cite{KNPR02,KN02} found a global sign mistake
in the GBE of the earlier work \cite{BPP96,HK98} which was confirmed by
their authors and by \cite{BCM02,RW02}.

Recently, Melnikov and Vainshtein pointed out new short-distance
constraints on the correlator (\ref{four}) \cite{MV04}, 
studied and extended in \cite{KPPR04}. 
The authors of \cite{MV04} constructed a model
which satisfies their main new short-distance
constraints and in which
the full hadronic light-by-light contribution 
is given by GBE and axial-vector
exchange contributions. 
Here we explicitly compare and comment on the
various contributions in the different calculations.

\section{``Old'' Calculations: 1995-2001}

With ``old'' we refer to the period 1995-2001. These calculations
were organized according to 
the large $N_c$ and CHPT countings discussed above \cite{EdR94}.
Notice that the CHPT counting was used just as a classification tool.
We want to emphasize once more that the calculations in
\cite{BPP96,HK98} showed that after several large
cancellations in the rest of the contributions,
 the numerically dominant one is the Goldstone boson
exchange.
 Here, we discuss mainly the work in \cite{BPP96},
with some comments and results from \cite{HK98} and \cite{KN02}.

\subsection{Pseudo-Scalar Exchange}

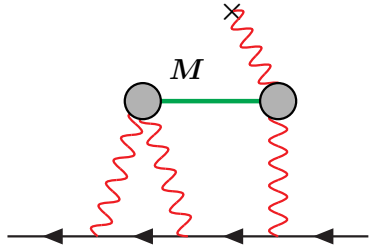
\begin{figure}
\begin{center}
\unitlength=1.7pt
\begin{picture}(80,80)
\SetScale{1.7}
\SetWidth{0.5}
\ArrowLine(20,0)(0,0)
\ArrowLine(40,0)(20,0)
\ArrowLine(60,0)(40,0)
\ArrowLine(80,0)(60,0)
\Text(50,50)[c]{\boldmath$\times$}
\SetColor{Red}
\Photon(20,0)(30,30){2}{6}
\Photon(40,0)(30,30){2}{6}
\Photon(60,0)(60,30){2}{6}
\Photon(60,30)(50,50){2}{5}
\SetColor{Green}
\SetWidth{1.}
\Line(30,30)(60,30)
\SetWidth{0.5}
\Text(40,35)[b]{\SetColor{Green}\boldmath$M$}
\SetColor{Black}
\GCirc(30,30){4}{0.7}
\GCirc(60,30){4}{0.7}
\end{picture}
\end{center}
\caption{A generic meson exchange contribution to the hadronic light-by-light
part of the muon $g-2$.}
\label{figexchange}
\end{figure}

The pseudo-scalar exchange  was saturated in \cite{BPP96,HK98}
by the Goldstone boson exchange.
This contribution is depicted in Fig.~\ref{figexchange}
with $M=\pi^0,\eta,\eta^\prime$.

Ref.\cite{BPP96},
used a variety of $\pi^0 \gamma^* \gamma^*$
form factors
\be
{\cal F}^{\mu\nu} (p_1,p_2) 
\equiv {N_c}/({6 \pi}) \,({\alpha}/{f_\pi})
\, i \, \varepsilon^{\mu\nu\alpha\beta} p_{1\alpha}
p_{2 \beta} \, {\cal F}(p_1^2, p_2^2)
\ee
fulfilling as many as possible QCD constraints.
A more extensive analysis of this form factor
was done \cite{BP01} finding very similar
numerical results.
In particular, the three-point form factors
${\cal F} (p_1^2,p_2^2)$ used in \cite{BPP96}
had the correct QCD short-distance behavior \footnote{
For this one and several other contributions \cite{BPP96} studied
thus the observance of QCD short-distance constraints,
contrary to the often stated claim that \cite{MV04}
is  the first
calculation to take such constraints into account, see e.g. \cite{ET06}.}
\ba
{\cal F} (Q^2,Q^2)  &\to& {A}/{Q^2} \, ,\quad
{\cal F} (Q^2,0)  \to {B}/{Q^2} \, ,
\ea
when $Q^2$ is Euclidean. These form factors were in agreement
with available data including the
slope at the origin as well as treating the  $\pi^0$, $\eta$ and $\eta'$
mixing.
All form factors converged for a cutoff scale 
$\mu \sim (2 - 4)$ GeV and produced  small  numerical differences
when plugged into the hadronic light-by-light contribution.

Somewhat different ${\cal F} (p_1^2,p_2^2)$ form factors
where used in \cite{HK98,KN02} but the results agree very well
(after correcting for the global sign).
For comparison, one can find 
the results of \cite{BPP96,HK98,KN02} in Tab.~\ref{tab1} 
 after  adding $\eta$ and $\eta'$ exchange contributions to the
dominant $\pi^0$ one.
\begin{table}
\begin{center}
\begin{tabular}{c|c}
 $\pi^0$, $\eta$ and $\eta'$ Exchange Contribution&
 $10^{10} \times a_\mu$\\
\hline
Bijnens, Pallante and Prades \cite{BPP96}  & 8.5 $\pm$ 1.3 \\
Hayakawa and  Kinoshita \cite{HK98}  & 8.3 $\pm$ 0.6 \\
Knecht and Nyffeler \cite{KN02}& 8.3 $\pm$ 1.2\\
Melnikov and Vainshtein \cite{MV04} & 11.4$\pm$1.0
\end{tabular}
\end{center}
\caption{Results for the $\pi^0$, $\eta$ and $\eta'$ exchange contributions.}
\label{tab1}
\end{table}

\subsection{Axial-Vector Exchange}

This contribution is depicted in Fig.~\ref{figexchange}
with $M=A=a_1^0,f_1$ and possible other axial-vector resonances.
For this contribution one needs the $A\gamma\gamma^*$ and
$A\gamma^*\gamma^*$ form factors.
Not much is known about these but there are  anomalous Ward identities
which relate them to the $P\gamma\gamma^*$ and $P\gamma^*\gamma^*$
form factors.

This contribution was not calculated in \cite{KN02}.
Refs. \cite{BPP96} and \cite{HK98} used nonet symmetry, 
which is exact in the large
$N_c$ limit, for the masses of the axial-vector resonances.
Their results are shown
in Tab.~\ref{tab2} for comparison.
\begin{table}
\begin{center}
\begin{tabular}{c|c}
 Axial-Vector Exchange Contributions
 & $10^{10} \times a_\mu$\\
\hline
Bijnens, Pallante and Prades \cite{BPP96}  & 0.25 $\pm$ 0.10 \\
Hayakawa and  Kinoshita \cite{HK98}  & 0.17 $\pm$ 0.10\\ 
Melnikov and Vainshtein \cite{MV04} & 2.2$\pm$0.5
\end{tabular}
\end{center}
\caption{ Results for the axial-vector exchange contributions.}
\label{tab2}
\end{table}

\subsection{Scalar Exchange}

This contribution is shown in Fig.~\ref{figexchange}
with $M=S=a_0,f_0$ and possible other scalar resonances.
For this contribution one needs the $S\gamma\gamma^*$ and
$S\gamma^*\gamma^*$ form factors. 
Within the extended Nambu--Jona-Lasinio (ENJL)
 model used in \cite{BPP96}, chiral 
Ward identities impose relations between the constituent
quark loop and scalar exchanges. The needed scalar form factors
are also constrained at low energies by CHPT.
Ref.~\cite{BPP96} used nonet symmetry for the masses.
This contribution was not included by \cite{HK98} and \cite{MV04}.

In \cite{BCM02} it was 
found the leading logarithms of the scalar exchange are the same as 
those of  the pion exchange but with opposite sign.
Ref.  \cite{BPP96} finds that sign for the full scalar exchange
contribution, obtaining
\be
a_\mu ({\rm Scalar}) = - (0.7\pm0.2) \cdot 10^{-10} \, .
\ee

\subsection{Other contributions at leading order in $1/N_c$.}

\begin{figure}
\label{quark}
\begin{center}
\epsfig{file=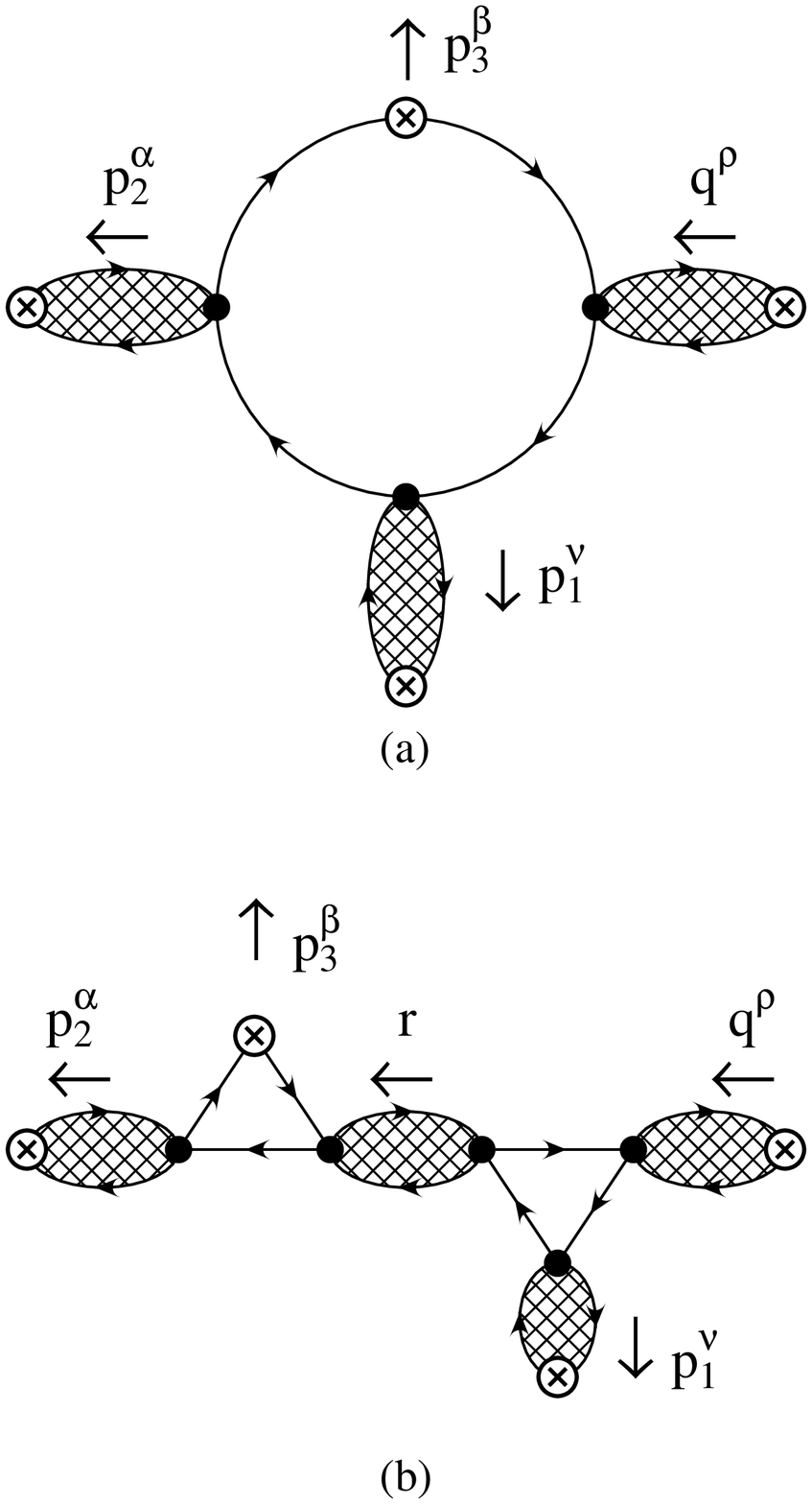,width=5cm,clip}
\end{center}
\caption{Quark-loop  contribution. as modeled in ENJL.}
\label{figquarkloop}
\end{figure}

This includes any other contributions that are not
exchange contributions. At short-distance, the main one is the quark-loop.
At long distances they are often modeled as a constituent quark-loop
with form-factors in the couplings to photons. 
This corresponds to the contribution 
shown in Fig.~\ref{figquarkloop}.
Ref.~\cite{BPP96} split up the  quark momentum integration into two pieces
 by introducing an Euclidean matching scale $\Lambda$. At energies
below $\Lambda$, the ENJL model was used
to compute the quark-loop contribution  while above $\Lambda$
 a bare (partonic) heavy  quark loop of mass
$\Lambda$ was used. The latter part scales as $1/\Lambda^2$ and
mimics the high energy behavior of QCD for a massless quark with
an IR cut-off  around $\Lambda$.  Adding these two contributions yields
 a stable result 
as can be seen in Tab.~\ref{quarkL}. 
\begin{table}
\begin{center}
\begin{tabular}{c|cccc}
$\Lambda$ [GeV] & 0.7 & 1.0 & 2.0 &4.0\\
\hline
\rule{0cm}{13pt} $10^{10} \times a_\mu$ & 2.2 &  2.0& 1.9& 2.0
\end{tabular}
\end{center}
\vspace*{0.2cm}
\caption{Sum of the short- and long-distance quark loop contributions
as a function of the matching scale $\Lambda$.}
\label{quarkL}
\end{table}

\subsection{NLO in $1/N_c$: Goldstone Boson Loops}

At next-to-leading order (NLO) 
in $1/N_c$, the leading contribution in the chiral counting 
to the correlator
in (\ref{Mlbl}), corresponds to charged pion and Kaon loops
which can be depicted analogously 
 to the quark-loop in Fig.~\ref{quark} but with pions and Kaons running
inside the loop instead.

In \cite{BPP96} the needed form-factors in the
$\gamma^*  P^+P^-$ and  $\gamma^* \gamma^* P^+P^-$  vertices
were studied extensively. In particular which forms were fully
compatible with chiral Ward identities were studied.
Full vector
meson dominance (VMD) is one model fulfilling the known constraints.
\footnote{Note that neither the ENJL model nor any fixed order in CHPT
was used.}
 The conclusion reached there was that there is a large ambiguity in the
momentum dependence starting at order $p^6$ in CHPT. Both 
the full VMD model
of \cite{BPP96} and the hidden gauge symmetry (HGS)
model of \cite{HK98} satisfy the known constraints.
Unfortunately, this ambiguity cannot be resolved
since there is no data for $\gamma^* \gamma^* \to \pi^+ \pi^-$.
Adding the charged pion and Kaon
loops, the results obtained in \cite{BPP96} 
and \cite{HK98} are listed in Tab.~\ref{tab3}.
\begin{table}
\begin{center}
\begin{tabular}{c|c}
 Charged Pion and Kaon Loop Contributions
 & $10^{10} \times a_\mu$\\
\hline
Bijnens, Pallante and Prades (Full VMD)
\cite{BPP96}  & $-$1.9 $\pm$ 0.5 \\
Hayakawa and  Kinoshita (HGS) \cite{HK98}  & $-$0.45 $\pm$ 0.85\\
 Melnikov and Vainshtein \cite{MV04} & 0$\pm$1.0 
\end{tabular}
\end{center}
\caption{ Results for the charged and Kaon loop contributions
to the hadronic light-by-light contribution to muon $g-2$.}
\label{tab3}
\end{table}

In view of the model dependence 
of this contribution,
the difference between \cite{BPP96}and \cite{HK98}
for this contribution needs to be added
 {\em linearly} to the  final  uncertainty of the hadronic
light-by-light contribution
to $a_\mu$.

\section{New Short-Distance Constraints: 2003-2004}

Melnikov and Vainshtein pointed out in \cite{MV04} a new
short-distance constraint on the correlator (\ref{four}).
This constraint is for
\ba
\langle T [ V^\nu(p_1) V^\alpha(p_2) V^\rho(-q=-p_1-p_2) ]| \gamma(p_3 \to 0)
\rangle
\ea
 and follows from the OPE for two vector currents
when $p_1^2\simeq p_2^2 >> q^2$:
\ba
T[ V^\nu(p_1) V^\alpha(p_2) ] \sim 
\varepsilon^{\nu\alpha \mu\beta} \, (\hat p_\mu/\hat p^2) \, 
[\overline q \hat Q^2 \gamma_\beta \gamma_5 q] (p_1+p_2)
\ea
with $\hat p =  (p_1-p_2)/2 \simeq p_1 \simeq -p_2$
and $\hat Q$ is the light quark electrical charge matrix
 (\ref{four}). This constraint was afterward
generalized in \cite{KPPR04}.

The authors of \cite{MV04}  saturated the full correlator
by exchanges. The new OPE constraint
is satisfied by introducing a pseudo-scalar exchange 
with the vertex on the $p_3$ side point-like.
This change strongly breaks the symmetry between the two ends
of the exchanged particle in Fig.~\ref{figexchange}.
Not all OPE constraints on the
correlator are satisfied at the same time  by this model,
but in \cite{MV04}
they argued that this made only a small numerical difference. 

To the pseudo-scalar exchange they added an axial-vector
exchange contribution which was found to be extremely sensitive
to the mixing of the resonances
 $f_1(1285)$ and $f_1(1420)$  as can be seen
in Tab.~\ref{massmixing}, taken from the results of \cite{MV04}.
\begin{table}
\begin{center}
\begin{tabular}{c|c}
Mass Mixing  & $10^{10} \times a_\mu$\\
\hline
No OPE and Nonet Symmetry with M=1.3 GeV& 0.3  \\
New OPE  and Nonet Symmetry with M= 1.3 GeV &  0.7   \\
New OPE  and Nonet Symmetry with M= M$_\rho$ &  2.8   \\
New OPE  and Ideal Mixing with Experimental Masses &  2.2 $\pm$ 0.5   
\end{tabular}
\end{center}
\vspace*{0.2cm}
\caption{Results quoted in \cite{MV04} for
the pseudo-vector exchange depending of the $f_1(1285)$ 
and $f_1(1420)$ resonances mass mixing.}
\label{massmixing}
\end{table}
The authors in \cite{MV04}  took the ideal mixing result for their
final result for $a_\mu$.

\section{Comparison}

Let us now try to compare the three calculations \cite{BPP96,HK98,MV04}.
In Tab.~\ref{comparison}, the results 
of the leading order in $1/N_c$ are shown.
The quark loop is of the same order  and has to be {\em added}
to get the full hadronic light-by-light while the model used
in \cite{MV04} is saturated just by exchanges.
\begin{table}
\label{comparison}
\begin{center}
\begin{tabular}{c|c}
Hadronic light-by-light 
at ${\cal O} (N_c)$  & $10^{10} \times a_\mu$\\
\hline
Nonet Symmetry + Scalar Exchange \cite{BPP96} & 
10.2 $\pm$ 1.9\\ 
Nonet Symmetry \cite{BPP96}&  10.9 $\pm$ 1.9  \\
Nonet Symmetry \cite{HK98} &  9.4 $\pm$ 1.6 \\ 
 New OPE and 
Nonet Symmetry \cite{MV04} &  12.1 $\pm$ 1.0 \\ 
 New OPE and 
Ideal Mixing  \cite{MV04} &  13.6 $\pm$ 1.5  
\end{tabular}
\end{center}
\vspace*{0.2cm}
\caption{Full hadronic light-by-light contribution
to $a_\mu$ at ${\cal O}(N_c)$. The difference between the
two results of \cite{BPP96} is the contribution of the
scalar exchange $-(0.7\pm0.1) \cdot 10^{-10}$.
This contribution was not included in
\cite{HK98} and \cite{MV04}.}
\label{largeN}
\end{table}
In the GBE the effect of the new OPE in \cite{MV04}
is a little larger than the quark loop
contributions of \cite{BPP96}. It also increases the axial-vector exchange
with nonet symmetry from 0.3 $\times 10^{-10}$ 
to 0.7 $\times 10^{-10}$
One thus sees a reasonable agreement in the comparison
of the  ${\cal O}(N_c)$
results of \cite{BPP96,HK98,MV04}
when using the  same mass mixing for the axial-vectors, namely, 
(10.9$\pm$ 1.9, 9.4 $\pm$ 1.6,  12.1 $\pm 1.0$). 

The final differences are due to the additional increase of
1.5$\times 10^{-10}$ from the ideal mixing in the axial vector exchange
in \cite{MV04} and the scalar exchange of $-$0.7$\times 10^{-10}$
in \cite{BPP96}.
 
Let us now see what are the different predictions at NLO in $1/N_c$.
In \cite{MV04}, the authors  studied the chiral expansion
of the charged pion  loop using the HGS model used in
\cite{HK98}. This model is known not to give the correct QCD high energy
behavior in some two-point functions, in particular it does not fulfill 
Weinberg Sum Rules,  see e.g. \cite{BPP96}.
Within this model, \cite{MV04} showed that there is 
a large cancellation between the first three terms 
of  an expansion of
the charged pion loop contribution
in powers of $(m_\pi/M_\rho)^2$.
It is not clear how one should interpret this. In \cite{BPP96}
some studies of the cut-off dependence of this contribution were done
and the bulk of their final number came from fairly low energies
which should be less model dependent.
However, it is clear that there is a large model dependence 
in the NLO in $1/N_c$ contributions.
But simply taking it to be
$(0\pm 1) \cdot 10^{-10}
$
as  in \cite{MV04} is rather drastic and certainly has
an underestimated error.

Let us now compare the results for the full hadronic 
light-by-light contribution to $a_\mu$ when summing all contributions.
The final result quoted in \cite{BPP96}, \cite{HK98} 
 and \cite{MV04} can be found  in Tab.~\ref{final}. The apparent agreement
between the  \cite{BPP96} and \cite{HK98} is hiding
non-negligible differences which numerically almost compensate.
There are differences in the quark loop and charged
pion and Kaon loops and \cite{HK98}does not include the scalar exchange.
\begin{table}
\begin{center}
\begin{tabular}{c|c}
Full Hadronic Light-by-Light & $10^{10} \times a_\mu$\\
\hline
Bijnens, Pallante and Prades
\cite{BPP96}&  8.3  $\pm$ 3.2\\ 
Hayakawa and Kinoshita \cite{HK98}&  8.9  $\pm$ 1.7\\ 
Melnikov and Vainshtein \cite{MV04}&  13.6  $\pm$ 2.5\\ 
\end{tabular}
\end{center}
\vspace*{0.2cm}
\caption{Results for the full hadronic light-by-light contribution
to $a_\mu$.} 
\label{final}
\end{table}

Comparing the results of \cite{BPP96} and \cite{MV04},
we have seen several differences of order
$1.5 \cdot 10^{-10}$, differences  which are not related
to the one induced by the new short-distance constraint
introduced in \cite{MV04}.
These differences are numerically of the same order or smaller than 
the uncertainty quoted in \cite{BPP96} but add up as follows.
The different axial-vector mass mixing account for $-1.5 \cdot 10^{-10}$,
the absence of scalar exchange in \cite{MV04} accounts
for $-0.7 \cdot 10^{-10}$ and the absence of the NLO in $1/N_c$
contribution in \cite{MV04} accounts for
$-1.9 \cdot 10^{-10}$. These model dependent
differences add up to   $-4.1 \cdot 10^{-10}$ 
out of the final $-5.3 \cdot 10^{-10}$
difference between the results in \cite{BPP96}
and \cite{MV04}.  Clearly, the new  OPE constraint
found in \cite{MV04} alone does not account for the large 
final difference as a reading of \cite{MV04} seems to suggest.

\section{Conclusions and Prospects}

 At present, the only possible conclusion is that the 
situation of the hadronic
light-by-light contribution to $a_\mu$ is unsatisfactory.
However, one finds a {\em numerical} agreement within roughly one sigma
when comparing the ${\cal O}(N_c)$ 
results found in \cite{BPP96,HK98,MV04}, 
see Tab.~\ref{largeN}. 
A new full ${\cal O}(N_c)$ calculation
studying the full correlator with the large
$N_c$ techniques developed in \cite{BGLP03,CEEKPP06}
and references therein, seems feasible and desirable. 

At NLO in $1/N_c$, one needs to control  
both Goldstone and non-Goldstone boson loop contributions.
The high model dependence of the Goldstone boson loop is clearly visible
in the different results of \cite{BPP96}
and \cite{HK98} and discussed in \cite{BPP96} and \cite{MV04}.
For non-Goldstone boson loops, little is known on how to 
consistently treat them, a recent attempt in another context is
\cite{RSP04}.

In the meanwhile, we propose as an educated guess\footnote{This educated guess 
 agrees with the one presented also at this meeting
by Eduardo de Rafael and by one of us, JB, at DESY Theory Workshop,
September 2005.}
\ba
\label{finalpluserror}
a_\mu= (11  \pm 4) \cdot 10^{-10} \, .
\ea
for the hadronic light-by-light contribution.
We believe that, that this number summarizes our present understanding
of the hadronic light-by-light contribution to $a_\mu$.
One can arrive at this number in several different ways:
the short-distance constraint and the ideal mixing for the
axial-vector exchange should lead to some increase of the results
of \cite{BPP96,HK98}; the scalar exchange and the pion and kaon loops
are expected to lead to some decrease of the result of \cite{MV04};
one can also average the leading in $1/N_c$ results (three middle results
of Tab.~\ref{largeN}). The final error remains a guess but the
error in (\ref{finalpluserror}) is chosen to include all the known
uncertainties.

\section*{Acknowledgments}

This work is supported in part by the European Commission (EC) RTN network,
Contract No.  MRTN-CT-2006-035482  (FLAVIAnet), 
the European Community-Research Infrastructure
Activity Contract No. RII3-CT-2004-506078 (HadronPhysics) (JB),
the Swedish Research Council (JB), 
 MEC (Spain) and  FEDER (EC) Grant No.
FPA2006-05294 (JP), and Junta
de Andaluc\'{\i}a Grant Nos. P05-FQM-101  and P05-FQM-437 (JP).

\end{document}